\RequirePackage{fix-cm}
\documentclass[smallextended]{svjour3}       
\smartqed  
\usepackage{url}
\usepackage{graphicx}
\usepackage{caption}
\usepackage{subcaption}
\usepackage{float}
\usepackage{glossaries}
\usepackage{multirow}
\usepackage{makecell}
\usepackage{booktabs}
\usepackage{multirow}
\usepackage{siunitx}
\usepackage{booktabs}
\usepackage{adjustbox}

\usepackage{caption}
\DeclareCaptionLabelSeparator{none}{ }
\usepackage[format=hang,font=small,labelfont=bf]{caption}

\makeglossaries

\newglossaryentry{lv}
{
    name=LV,
    description={Left Ventricle}
}
\DeclareUnicodeCharacter{2009}{\,} 
\newglossaryentry{MI}
{
    name=MI,
    description={Myocardial Infarction}
}
\newcounter{alphasect}
\def\alphainsection{0}

\let\oldsection=\section
\def\section{%
  \ifnum\alphainsection=1%
    \addtocounter{alphasect}{1}
  \fi%
\oldsection}%

\renewcommand\thesection{%
  \ifnum\alphainsection=1%
    \Alph{alphasect}
  \else%
    \arabic{section}
  \fi%
}%

\begin{document}

\title{Fully Automated 2D and 3D Convolutional Neural Networks Pipeline for Video Segmentation and Myocardial Infarction Detection in Echocardiography
}
\subtitle{}


\author{Oumaima Hamila \and Sheela Ramanna \and Christopher J. Henry \and Serkan Kiranyaz \and Ridha Hamila \and Rashid Mazhar \and Tahir Hamid }


\institute{Oumaima Hamila, Sheela Ramanna, Christopher J. Henry \at
              Department of Applied Computer Science, The University of Winnipeg\\
              515 Portage Avenue, Winnipeg, MB Canada, R3B 2E9 \\
              \email{hamila-o@webmail.uwinnipeg.ca}, {\{s.ramanna,ch.henry\}@uwinnipeg.ca}         
			\and			
		   Serkan Kiranyaz, Ridha Hamila \at
           Department of Electrical Engineering, Qatar University, Doha, Qatar \\
           \email{\{mkiranyaz,hamila\}@qu.edu.qa}           
           \and           
           Dr. Rashid Mazhar \at
           Thoracic surgery, Hamad hospital, Hamad Medical Corporation, Qatar \\
           \email {rashmazhar@hotmail.com}                     
           \and           
           Dr. Tahir Hamid \at
           Cardiology, Heart hospital Hamad Medical Corporation, Qatar \\
           \email {tahirhamid76@yahoo.co.uk}
           }

\date{Received: date / Accepted: date}
\authorrunning{}
\titlerunning{ }

\maketitle

\begin{abstract}
Myocardial infarction (MI) is a life-threatening disorder that occurs due to a prolonged limitation of blood supply to the heart muscles, and which requires an immediate diagnosis to prevent death. To detect MI, cardiologists utilize in particular echocardiography, which is a non-invasive cardiac imaging that generates real-time visualization of the heart chambers and the motion of the heart walls. These videos enable cardiologists to identify almost immediately regional wall motion abnormalities (RWMA) of the left ventricle (LV) chamber, which are highly correlated with MI. However, data acquisition is usually performed during emergency which results in poor-quality and noisy data that can affect the accuracy of the diagnosis. To address the identified problems, we propose in this paper an innovative, real-time and fully automated model based on convolutional neural networks (CNN) to early detect MI in a patient’s echocardiography. Our model is a pipeline consisting of a 2D CNN that performs data preprocessing by segmenting the LV chamber from the apical four-chamber (A4C) view, followed by a 3D CNN that performs a binary classification to detect MI. The pipeline was trained and tested on the HMC-QU dataset consisting of 162 echocardiography. The 2D CNN achieved 97.18\% accuracy on data segmentation, and the 3D CNN achieved 90.9\% accuracy, 100\% precision, 95\% recall, and 97.2\% F1 score.  Our detection results outperformed existing state-of-the-art models that were tested on the HMC-QU dataset for MI detection. This work demonstrates that developing a fully automated system for LV segmentation and MI detection is efficient and propitious.

\end{abstract}

\section{Introduction}
\indent 

An early detection of myocardial infarction (MI) \cite{MIdef}, which is medically referred to as heart attack, can prevent a patient from enduring several health complications such as heart failure or sudden death \cite{PMID,leading}. In fact, MI is pathologically defined as the death of the myocardial cells due to extended cardiac ischemia, that is defined as an abrupt and prolonged limitation of blood supply to the heart muscles \cite{test}. Once a patient is suspected with MI, an immediate and accurate diagnosis should be performed to detect early and basic symptoms associated with the disease, which are defined predominantly as an abnormal or a non-uniform motion, known as hypokinesia, of one or several regional sections of the left ventricle (LV) wall of the heart \cite{LVfunctionInMI}. As soon as regional wall motion abnormalities (RWMA) of the LV are perceived \cite{Adrian,Vittorio}, the process of infarction can be completely aborted within the first hour \cite{Roberts} and the affected patient can avoid serious or fatal health complications \cite{death}. However, the major constraints to an accurate diagnosis are the unavailability of an end-to-end, rapid and exact assessment tool to reliably detect MI in real-time. 
\newline \indent 
During the last two decades, non-invasive imaging for cardiovascular disease (CVD) diagnosis and monitoring have witnessed an important evolution \cite{DW,AdvancedImaging,future0of0cardiac0imaging}, which enabled cardiologists to further develop their understanding of cardiac pathologies and, in particular, benefitted MI analysis, identification and treatment \cite{personal}. To detect and assess RWMA of the cardiac chambers, echocardiography is highly recommended by The American Society of Echocardiography because of its capability to assess in real-time both the cardiac function and structure \cite{asc}. It generates important amounts of visual data including the size and shape of the heart and its chambers and the motion of the heart walls while they are beating. This helps cardiologists to identify RWMA in a patient’s echocardiography and assign the adequate treatment immediately \cite{immediate}, which may minimize the damage on the cardiac muscle tissues and prevent patients from facing death \cite{qatar3}. 
\newline \indent
Thanks to the evolution of modern computing technologies, such as machine learning (ML) and deep learning (DL), and the advent of supercomputers \cite{AI,DLandBiomed,DLinRadiology,potential}, it is now possible to learn optimized patterns that are highly accurate from large volumes of data with minimal human intervention. Some works in early detection of MI in echocardiography used ML and convolutional neural networks (CNN) \cite{deepLearningMRI,machineLearning2Decho,qatar2}, while some others were based on classical approaches such as Metaheuristics \cite{geneticAlgorithm} and Fourier tracking \cite{FourierTracking}. Some methods either heavily rely on very specific and limited conditions of data acquisition (high-resolution echocardiograms, high frame-rate, minimal noise) \cite{AIearlyPrediction}, that require the technician or the cardiologist to perform preliminary preprocessing steps to be able to proceed with the prediction process \cite{ANNclassifyLV}, or extract finite and restricted features to be used with the classification model. 
\newline \indent
Even though echocardiography is an ideal tool to detect RWMA during a myocardial ischemia, some cardiologists find it challenging to use as a primary diagnosis tool and often employ other diagnosis methods to determine the disease, such as electrocardiogram or angiogram \cite{circa}, due to their straightforwardness or simplicity to use and interpret. For instance, echocardiography produces large and complex data that needs to be entirely exploited and understood in order to make a complete diagnosis based on visual interpretation \cite{AI}, which is highly dependent on the level of experience of the cardiologist in question \cite{QuantitaiveDetection}. Moreover, in some cases, an important amount of the generated data remains unused due to insufficient time and difficulty in interpretation \cite{standardTTE}. Furthermore, data acquisition is usually performed in emergencies, which often yields images of low quality \cite{contrast,emergencyECHO}, that may also be of a low-resolution because of the cardiac machine characteristics itself. As a result, these constraints negatively impact the accuracy of the MI diagnosis \cite{ImpactOI}. Thus, there is a need to create an advanced, reliable and fully-automated process that efficiently uses echocardiography to perform accurate MI detection in real-time \cite{BayesianNetworks,AIearlyPrediction}.
\newline \indent
In this paper, we propose a novel method to overcome the following issues: i) subjective reading of the data that relies on expert cardiologists, ii) generated poor-quality and low-resolution echocardiography, iii) massive amounts of video data that requires preprocessing prior to detection, and iv) slow, manual and inefficient MI detection. The proposed solution is an end-to-end and fully automated pipeline consisting of a 2D CNN that performs data preprocessing followed by a 3D CNN that performs binary classification to detect MI from an echocardiography in real-time. The pipeline begins with a 2D CNN that segments the LV region from an echocardiography, since the occurrence of MI is highly correlated with RWMA of the LV walls \cite{LVfunctionInMI}. Then, the segmented video is fed to a 3D CNN, which extracts from it the relevant spatio-temporal features and uses them to detect MI. The input of the pipeline is an unprocessed echocardiography of a patient as acquired by a technician or a cardiologist, and the output is the detection result, which is either abnormal (MI) or normal (N). Both 2D and 3D CNNs were trained and tested on the HMC-QU benchmark dataset \cite{HMC-QU}, which contains 162 4-chamber view echocardiography recordings obtained at the Hamad Medical Corporation (HMC) \cite{hamadMed} between 2018 and 2019 and approved for scientific use in February 2019. The echocardiography videos represent 93 patients that were diagnosed with MI, while the remaining 69 videos represent normal patients. 
\newline \indent
The main contributions of this work are:
\begin{description}
\item[$\bullet$] A fully automated pipeline for video segmentation and MI detection in echocardiography.
\item[$\bullet$] An indiscriminative pipeline that processes videos of different sizes, different frame rates and different resolutions.
\item[$\bullet$] An early and real-time MI detection model in echocardiography.
\item[$\bullet$] A robust pipeline that performs fairly well on low-quality videos corrupted with intense noise.
\item[$\bullet$] A system for LV segmentation in echocardiography that achieved 97.18\% accuracy on the HMC-QU benchmark dataset.
\item[$\bullet$] A system for MI early detection in echocardiography that achieved 90.9\% accuracy, 100\% precision and 95\% recall on the HMC-QU benchmark dataset. 
\item[$\bullet$] A lightweight system that runs on parallel threads and does not require high memory or computational power in order to be executed, which makes the system adequate to be embedded in external devices.
\item[$\bullet$] Compared with two state-of-the-art methods that performed MI detection on the HMC-QU benchmark dataset, our system achieved higher results.
\end{description}

The outline of this paper is as follows. In Section 2, we discuss the main advantages and drawbacks of state-of-the-art MI detection methods. We then explain in Section 3 the pipeline architecture and discuss details related to the dataset. In Section 4, we explain the preprocessing techniques applied to the dataset which is used as input to a 2D CNN. In addition, we present the details related to the 2D CNN architecture. We describe data preprocessing techniques applied to the processed videos before feeding it to the 3D CNN in Section 5, together with the details of the 3D CNN architecture. In Section 6, we describe the training processes and the evaluation metrics related to each model, followed by a discussion of the results. Finally, in Section 7, we present concluding remarks.

\section{Related Work}
\label{relatedWork}
\indent

Cardiac imaging technologies have been evolving during the last few decades into more advanced machines that generate complex and detailed data, such as real-time videos of the hearts’ chambers and valves, which have inspired scientists along with cardiologists to develop newer methods that aim to detect and assess cardiac deficiencies based on evaluating these data.
Therefore, multiple techniques, which have been produced over the years to detect cardiac diseases by evaluating the myocardial motion, are based either on signal-processing, chemical-processing, image-processing or, more recently, video-processing. 
\newline \indent 
In \cite{dlv}, a contour-based technique for detecting wall motion abnormality by analyzing the temporal pattern of normalized wall thickening was proposed. Epicardium and endocardium zones were manually extracted by segmenting images representing 27 real-life patients. Subsequently, AHA 17-segment model was used to evaluate regional wall changes in normalized wall thickness followed by a Naïve Bayes classifier. Although the model achieved 100\% true-negative, it only obtained 70\% true-positive for apical 4-chamber (A4C) view, which means that the model does not predict the cardiac disease when it happens 30\% of the time. Moreover, manual preprocessing is time-consuming, subject to human error and relies on human expertise to perform the segmentation task properly, which may affect the accuracy and the quality of the results. 
\newline \indent
In \cite{quant}, existing quantitative approaches were applied to detect and identify localized wall motion abnormalities from 12-lead ECG, 2D echocardiography images and coronary angiography in patients affected with MI. Adequate 2D echocardiography images representing 4 different cardiac views were obtained from 74\% of well-defined patients, and used to assess abnormal segments, which are characterized by a standard deviation that is inferior to the average contraction estimated over 10 normal subjects. ECG and angiography data were analyzed independently by two observers. The results concluded that 2D echocardiography images allow an extended assessment of endocardial wall motion and regional wall thickening and can be applied for a quantitative approach to detect regional LV abnormalities, while ECG and angiography have certain limitations. Furthermore, results showed that area methods performed better than linear methods by achieving at best 95\% of accuracy versus 84\% for linear models in predicting localized regional LV contraction deficiencies. Even though the accuracy of the prediction is 95\%, the model does not allow an early detection of LV abnormalities. In fact, it only examines the effects of CAD on myocardial performance. 
\newline \indent
Another approach to assessing data generated from cardiac imaging uses ML and DL models. In \cite{1year}, 723,754 clinically acquired echocardiographic videos of the heart (~45 million images) of 27,028 patients were evaluated to predict 1-year mortality rate in patients who had encountered heart deficiencies. The dataset was divided into 21 groups such that each group represented a standard echocardiographic view. Then, distinct 3D CNN models were generated, trained and tested on each data group separately. Additionally, longitudinal electronic health records were added to the videos as input data to the models during training, and the accuracy of the 1-year mortality prediction in patients with heart abnormality records was 75\%. This paper shows that applying 3D CNN to echocardiography videos to perform a prediction task is efficient and plausible, however, there is still room to improve the accuracy. For example, adding a preprocessing model to normalize the raw videos prior to training the CNN models could ameliorate the accuracy of the prediction.
\newline \indent
The authors in \cite{assess} used DL in order to assess regional wall motion abnormality in Echocardiographic images. Data from 300 patients with a history of MI were divided into 3 groups such that each data group contained images representing a distinct cardiac abnormality. Data from 100 healthy patients were also included as a $4^{th}$ data group. Then, only images with good or adequate acoustic detail were selected while poor quality images were discarded from the final dataset. Images were then standardized to the same spatial dimensions and fed to 10 versions of the same DCNN model to detect the presence of RWMA. In comparison to the prediction outcome performed by two expert cardiologists, the DCNN produced similar results such that the AUC curve produced by the cardiologists was similar to that produced by the DCNN (0.99 vs 0.98). Even though the prediction results are considerably fair, the model was trained only on good-quality echocardiographic images, which implies that testing it over real-life images that potentially contain noise, missed information and inadequate acoustic detail could reduce its performance and may lead to erroneous predictions. 
\newline \indent
In \cite{serum}, both electrocardiogram and serum analysis were used to detect acute MI in 82 patients who were suspected of having MI within one hour of their arrival to the care unit. The electrical activity of the heart produced by the 12-lead electrocardiogram was recorded and a 10 ml blood sample was obtained within the first hour of their admission to the care unit, and all the data was analyzed by two observers. Moreover, several chemical substances such as creatine kinase and myoglobin were measured, which takes 10 minutes to perform. These parameters were combined to perform a logistic regression analysis that led to the detection of MI by 64\% of accuracy. The model mandatorily needs a chemical examination that is source-intensive in order to make an assessment that is only 64\% accurate, which may not be sufficient for reliable MI detection.
\newline \indent
A more recent study \cite{qatar3} developed an MI detection model based on local motion estimation in an attempt to overcome the limitations of speckle tracking methods, which achieved at their best around 85\% of sensitivity and specificity, by investigating more reliable and robust wall motion analysis models. Although both speckle tracking and local motion estimation use a standard LV segmentation model from which they analyze the motion of each segment of the LV wall separately to determine signs of RWMA and therefore detect MI, speckle tracking relies solely on assessing the motion of a single speckle per segment, while local motion estimation uses several speckles per segment. However, the accuracy of both techniques can be negatively impacted by noisy echocardiography, since noise makes segment’s tracking difficult and unreliable. Therefore, the authors further explored the use of local motion estimation by developing an approach based on Active Polynomial which captures the LV contour and divides it into 7 segments. The motion of each segment was then evaluated to detect RWMA and MI was predicted if at least one LV segment motion was determined to be abnormal. The model’s performance metrics were estimated on two subsets on the HMC-QU benchmark dataset; the first subset contained the totality of the videos while the second subset contained only reasonable quality echocardiography videos, which were selected based on a visual assessment by the authors. The reported results showed that the model achieved better MI detection accuracy on reasonable quality videos than on the totality of the dataset, by achieving 87\% accuracy on the second subset and only 83\% on the first. Despite attaining better results than current state-of-the-art methods, Active Polynomials performance decreases considerably when tested on low-quality or low-resolution echocardiography videos, which negatively impacts the efficiency and robustness of the method. 
\newline \indent
In another recent study \cite{qatar1} where the HMC-QU benchmark dataset was used, a three-phase early MI detection approach for low-quality echocardiography was developed. The model’s architecture begins with an encoder-decoder CNN (E-D CNN) for LV wall segmentation. The E-D CNN was followed by a feature engineering method that analyzed 5 and 6 segments from the segmented LV wall to extract two motion-based features and one area-based feature, that were used to create two datasets  based on which were trained four conventional ML models for early MI detection in the third phase. The reported performance metrics showed that detecting MI from either 5-segment features data or 6-segment features data gave almost the same accuracy and precision. Furthermore, all four ML models achieved nearly the same results, and overall, the best achieved accuracy, sensitivity and specificity are 80\%, 85\% and 74\% respectively. Although the model’s main contribution is an operator-independent MI assessment that provides quantitative measurements for the LV wall segments, the attained evaluation metrics can be further improved possibly by a more extended feature extraction method and a more accurate classification. 
\newline \indent
Although all aforementioned studies have explored a variety of approaches to develop a robust and efficient MI detection models from echocardiography videos and have succeeded to a certain extent in obtaining fairly satisfactory results, the potential for improvement to attain higher and more robust performances still exists. While knowing that any non-zero false-negative rate can cause the loss of multiple human lives due to wrong diagnosis and that late MI detections can delay patients’ treatments and cause irreversible damage on their health, developing more precise and robust models remains necessary. Our approach aims to overcome most of the drawbacks observed and learnt from most recent MI detection research works. Thus, we created an end-to-end fully automated pipeline which leaves no room for manual processing that may lead to generating unreliable, erroneous or subjective segmentation or detection results. Furthermore, we used the totality of the HMC-QU benchmark dataset which mainly consists of low-quality and low-resolution echocardiography videos, and we applied a 2D CNN for LV segmentation then used a 3D CNN as a feature extractor and a densely connected neural network as a classifier to achieve the highest MI detection accuracy regardless of the video quality in order to create a robust diagnosis approach. Moreover, our approach is lightweight, runs in real-time and on parallel threads, which accelerates the diagnosis results especially in case of emergencies.

\section{Methodology}
\label{methodology}
\indent

The main goal of our work is to create a fully automated system for LV segmentation and MI detection in echocardiography in order to assist technicians and cardiologists in the process of diagnosing a patient affected or suspected with MI. This system must be robust, and as accurate and efficient as possible. In emergencies, for example, echocardiography acquisition tends to be made hastily, which may negatively impact the video’s quality and content. Moreover, most of the employed echocardiogram machines produce videos of low-resolution and which their frame rate is generally below 30 frames per second (fps). In the following sections, we give an overview of the pipeline architecture and a description of the echocardiography videos acquired for this work.
\subsection{Pipeline Overview}

\begin{figure*}[h!]
\centering
  \includegraphics[scale=0.8]{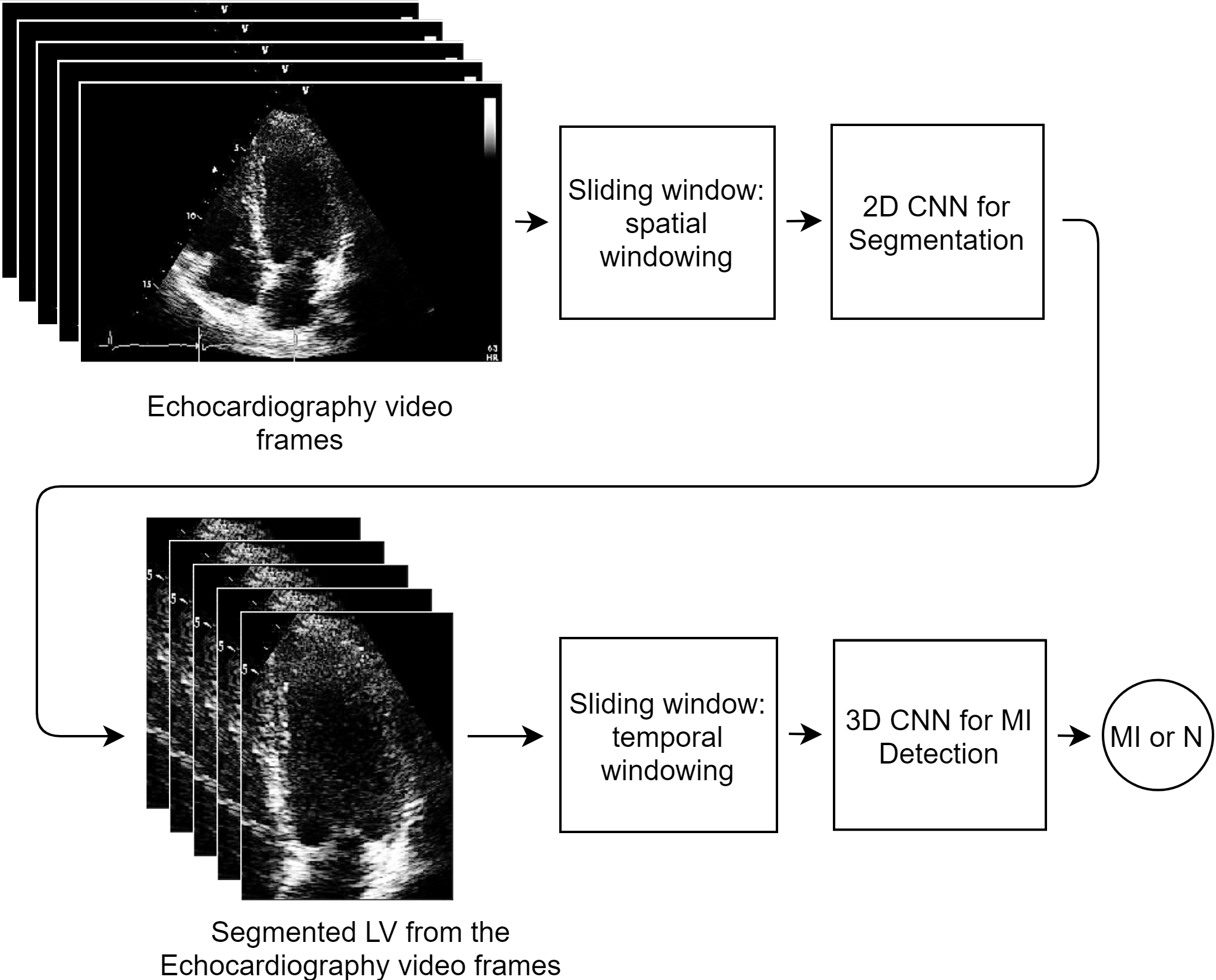}
\caption{Fully automated pipeline for MI detection, where the input is an echocardiography video and the output is the prediction results}
\label{fig:pipeline}
\end{figure*}

Figure \ref{fig:pipeline} illustrates the flow of the automated pipeline where the input consists of \textit{echocardiography video frames}, and the output is the MI detection result, \textit{abnormal (MI) or normal (N)}. The echocardiography frames are processed by the sliding window technique which divides each frame into spatial windows of equal dimension. The spatial windows are passed through the 2D CNN to segment the LV from each frame’s spatial windows. Once the segmented windows are produced, they are reassembled into segmented frames. These segmented frames are reassembled to produce a segmented video, where the order of appearance of each segmented frame is kept in the same order of appearance in the original echocardiography video. The segmented video frames are labelled in Figure \ref{fig:pipeline} as the \textit{segmented LV from the echocardiography video frames}. These are then processed by another sliding window to produce temporal windows of the same dimensions. The temporal windows are then passed through a 3D CNN that classifies them into one of the two classes: MI or N . The final class of the input video is estimated as the statistical mode of all the predictions of the frames constituting the video. 

\subsection{HMC-QU Dataset}
\indent 

\begin{figure}[h!]
\centering
  \includegraphics[width=1\linewidth]{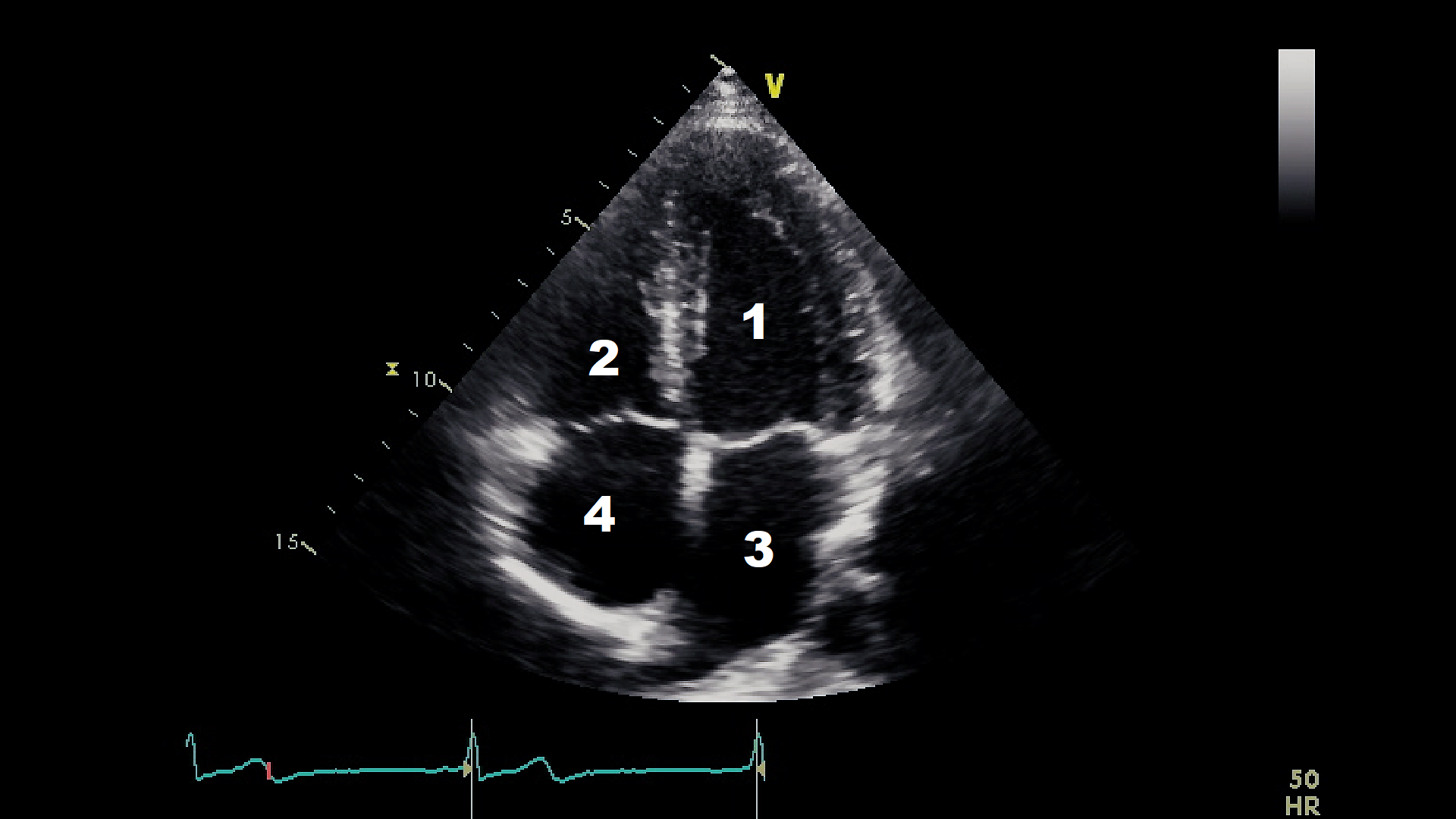}
\caption{Apical four-chamber view. The numbers from 1 to 4 marking the four different chambers correspond respectively to the LV, the right ventricle, the left atrium, and the right atrium}
\label{fig:a4c}
\end{figure}

In collaboration with our cardiologist co-authors practising at HMC Hospital and Qatar University, we have used a dataset of 162 annotated echocardiography videos from the study number MRC-01-19-052 approved by the Medical Research Centre at HMC. The dataset contains 162 A4C view recordings of 2D echocardiography, obtained between 2018 and 2019, and were approved for scientific use by the local ethics board of HMC in February 2019. The echocardiography videos have a frame rate of 25 fps and their spatial resolution varies between $422\times636$ pixels (px) and $768\times1024$ px. Furthermore, several echocardiography videos have corrupt content due to either noise or distorted representation of the A4C view, which usually consists of missing parts of the heart chambers that could not be acquired during acquisition. In this work, we used the entire HMC-QU dataset which included videos with both reasonable quality and poor quality. The dataset became publicly available in February 2021.
\newline\indent
This work focuses on performing early MI detection by learning RWMA of the LV wall in the A4C view echocardiography. Figure \ref{fig:a4c} shows a captured frame representing the A4C view, which consists of four distinct heart chambers, numbered from 1 to 4, where 1 identifies the LV, 2 to 4 identifies the Right Ventricle, the Left Atrium, and the Right Atrium, respectively. 

\begin{figure}[h!]
    \centering
    \begin{subfigure}[b]{0.49\textwidth}
        \includegraphics[width=1\linewidth]{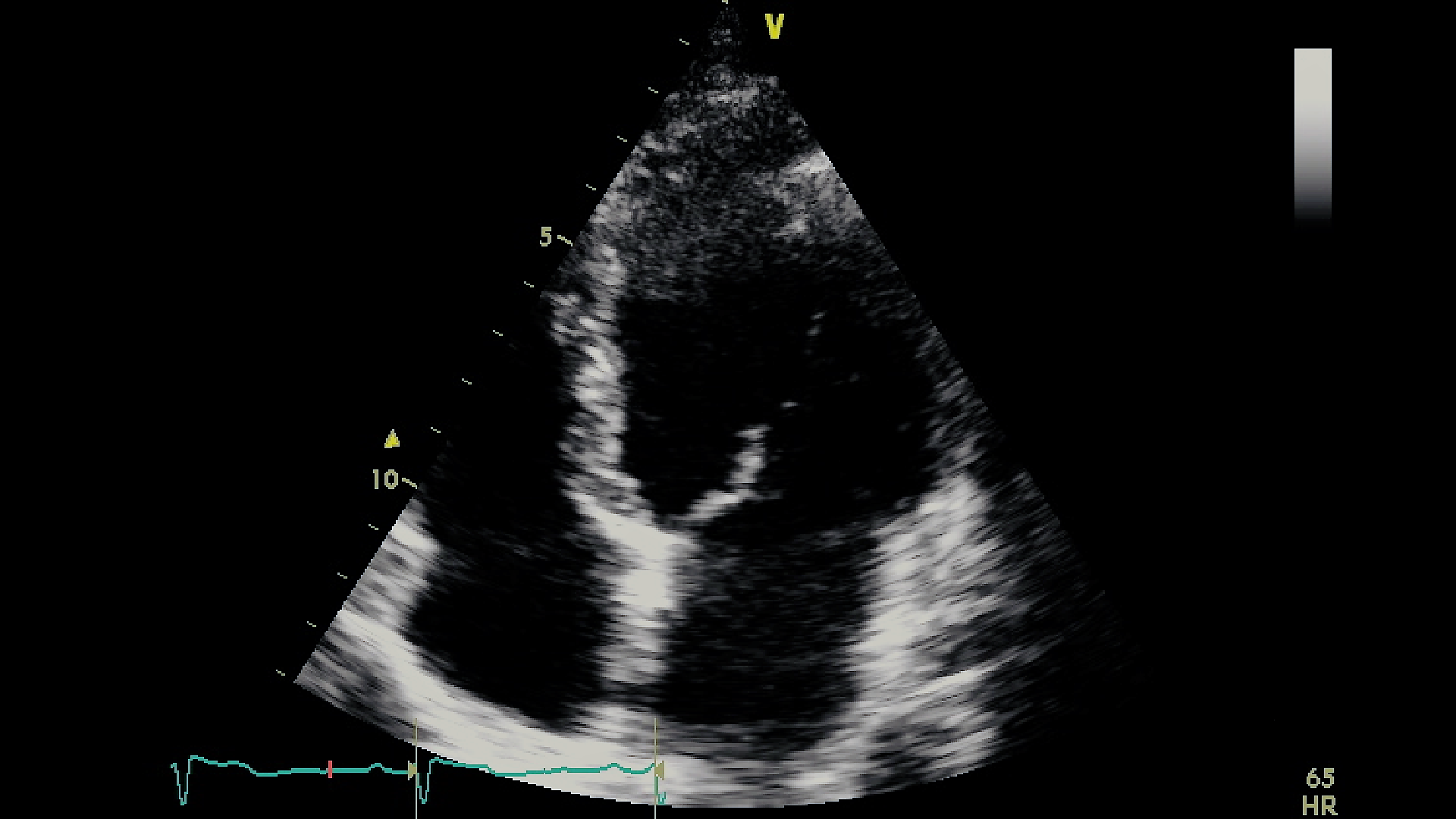}
        \caption{}
        \label{quality_a}
    \end{subfigure}
    \begin{subfigure}[b]{0.49\textwidth}
        \includegraphics[width=1\linewidth]{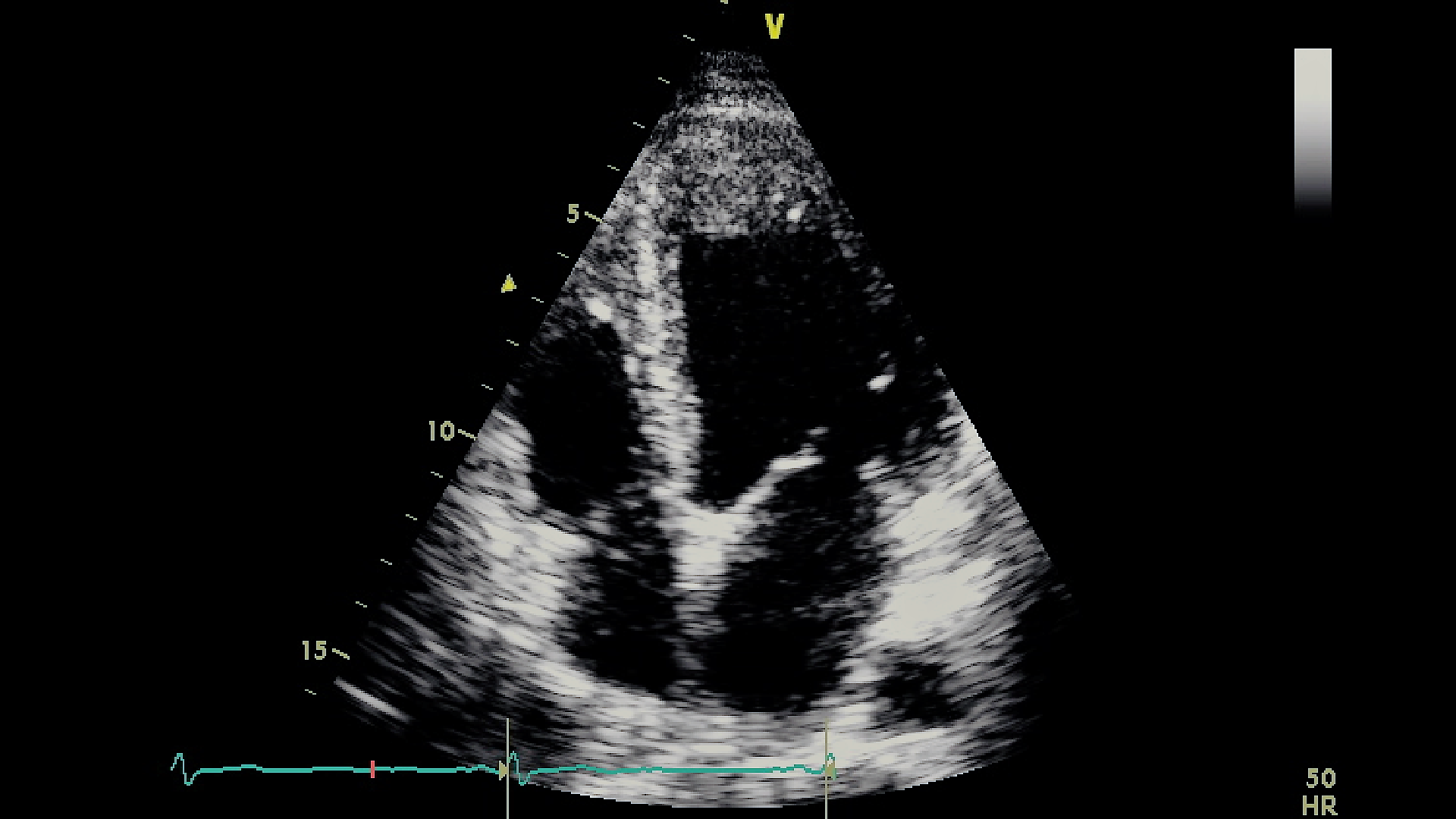}
        \caption{}
        \label{quality_b}
    \end{subfigure}
    \begin{subfigure}[b]{0.49\textwidth}
        \includegraphics[width=1\linewidth]{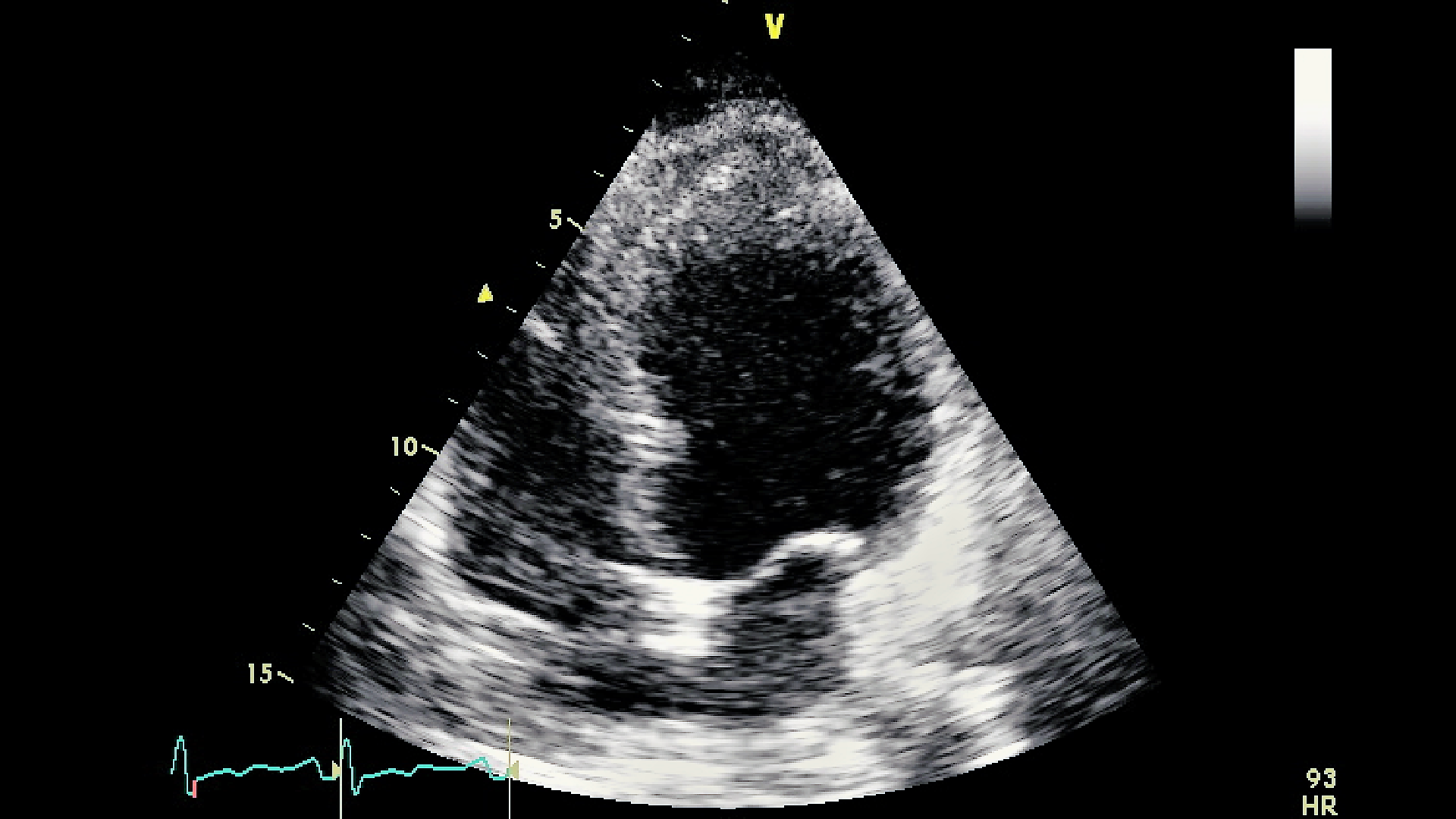}
        \caption{}
        \label{quality_c}
    \end{subfigure}
    \begin{subfigure}[b]{0.49\textwidth}
        \includegraphics[width=1\linewidth]{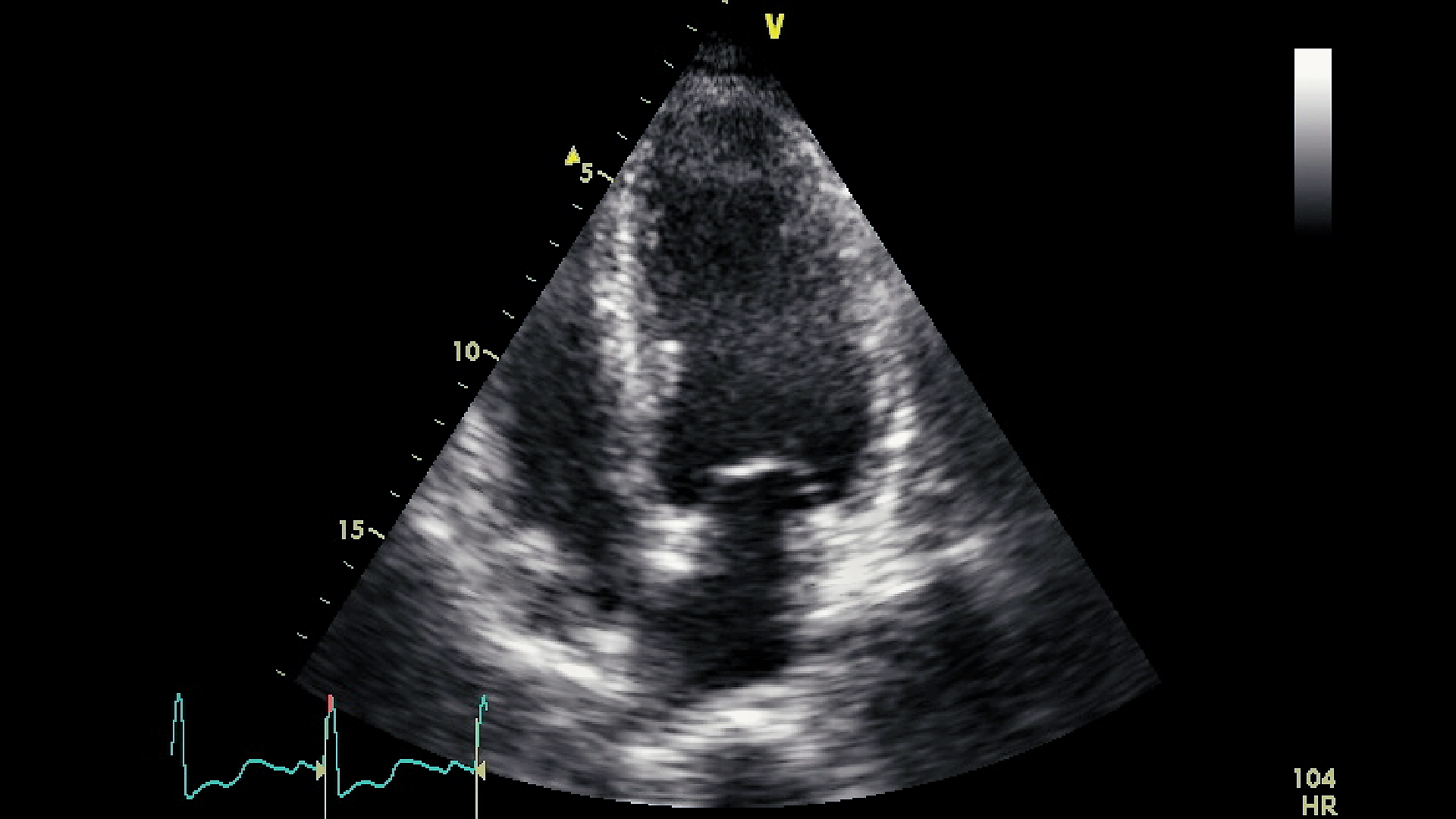}
        \caption{}
        \label{quality_d}
    \end{subfigure}
    \begin{subfigure}[b]{0.49\textwidth}
        \includegraphics[width=1\linewidth]{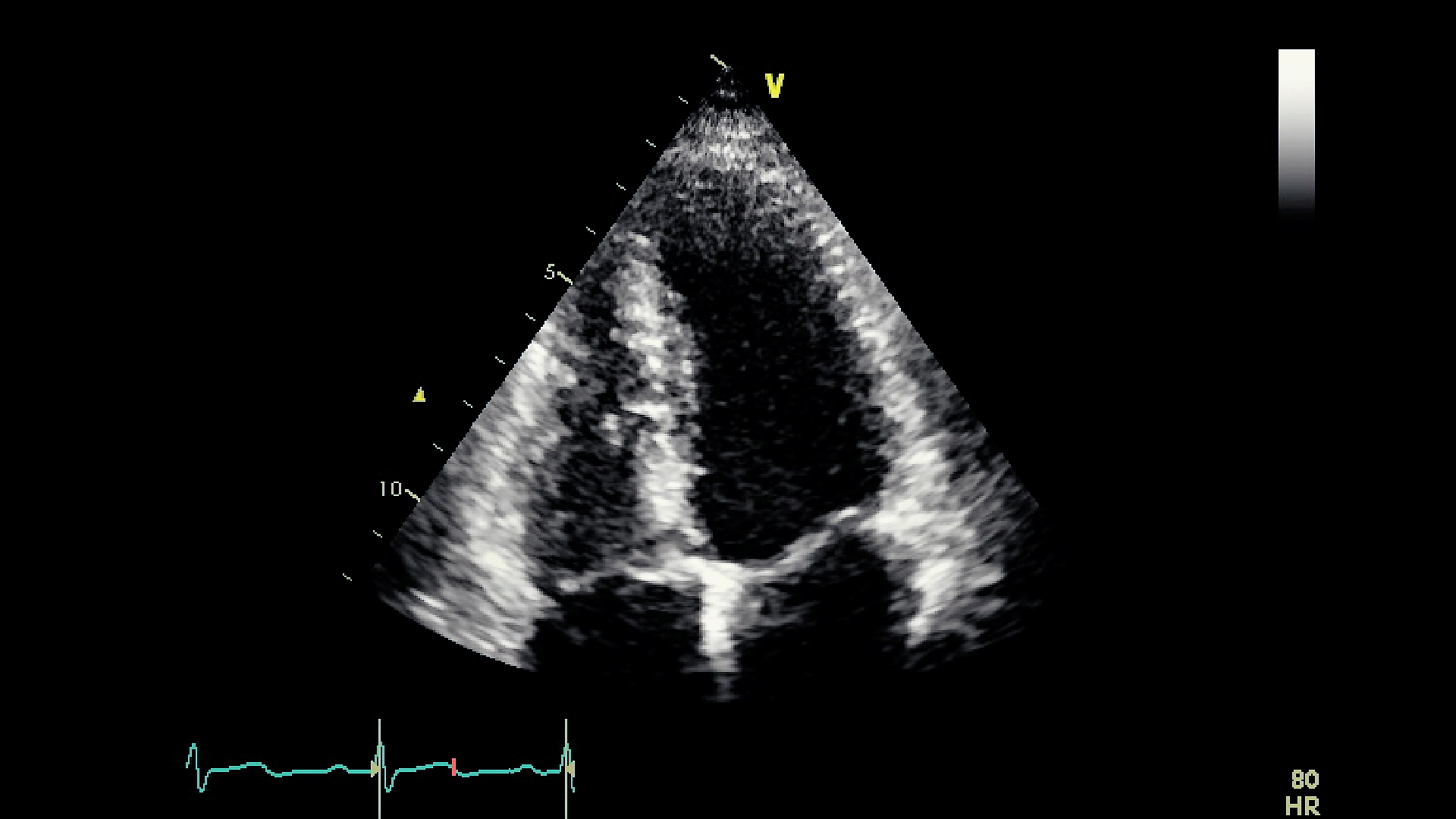}
        \caption{}
        \label{quality_e}
    \end{subfigure}
    \begin{subfigure}[b]{0.49\textwidth}
        \includegraphics[width=1\linewidth]{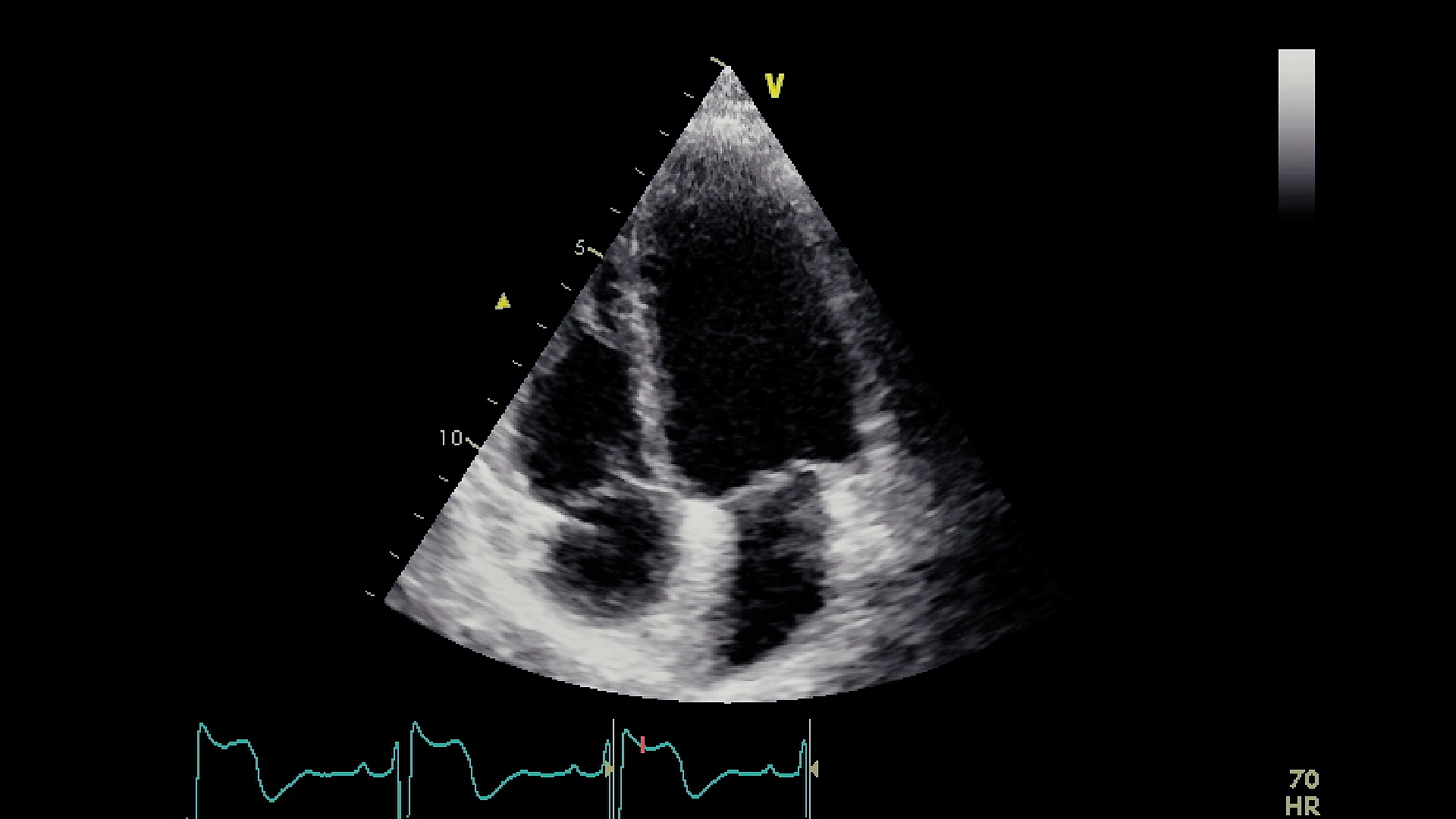}
        \caption{}
        \label{quality_f}
    \end{subfigure}
    \caption[]{Captured frames from 6 different videos of our dataset, where each image from \ref{quality_a} to \ref{quality_f} corresponds to a distinct video. \ref{quality_a} represents a blurred left wall in the LV, \ref{quality_b} represents a missing left wall in the LV, \ref{quality_c} represents blurred LV walls, \ref{quality_d} represents noise inside the LV, and \ref{quality_e} and \ref{quality_f} represent normal echocardiograms}
    \label{fig:examples}
\end{figure}

Figure \ref{fig:examples} represents captured frames representing the quality of several videos from our dataset, which varies from good to noisy. Figures from \ref{quality_a} to \ref{quality_f} correspond to distinct frames each captured from different videos. We notice that in Figure \ref{quality_a} the left wall of the LV is blurred. Also, in Figure \ref{quality_b}, the left wall of the LV is blurred and almost missing. In the same way, we observe that the totality of the LV wall is blurred in Figure \ref{quality_c}; and that the interior of the LV is disrupted with noise in Figure \ref{quality_d}. Finally, both Figure \ref{quality_e} and Figure \ref{quality_f} show acceptable LV representations, where the LV walls are captured and the chamber's interior is empty from noise. Moreover, since our study is centered on the LV chamber only, we purposely ignore the distortions of the rest of the cardiac chambers (Right Ventricle, Left Atrium, and Right Atrium) in the dataset videos. For example, in Figure \ref{quality_e}, both the Left Atrium and the Right Atrium are partially cut from the view, however, this does not impact our study.

Hence, our final set of videos for segmentation consists of both clear and blurred video images of the LV chamber.
\section{Video Segmentation with 2D CNN}
\label{segmentaionSection}
\indent

The 2D CNN performs a supervised classification by learning to map the input echocardiography video to its adequate segmentation mask. Thus, we manually created segmentation masks that cover the LV chamber from the A4C view and discards the remaining chambers. The manually created segmentation masks were assigned to the dataset video frames as labels, and fed to the 2D CNN to learn the best  segmentation mask from any given echocardiography video. The videos were normalized prior to training the 2D CNN by means of the sliding window technique due to differences in the dimensions of the frames.

\subsection{Data Preprocessing for 2D CNN}
\subsubsection{Creating labels}
\indent

The first step was preparing a labeled dataset, where each input is an echocardiography video frame, and each output is a corresponding segmentation mask. The segmentation masks were manually created and designed to cover the area of LV from the A4C in all the frames included in a given video. In each video, at least one cardiac cycle was performed, which means that we have at least one diastole (when the heart refills with blood) and one systole (when the heart pumps the blood) per video. The segmentation mask boundaries were determined such that they form a rectangle that encompasses the totality of the LV even on the frames where the heart is fully expanded, i.e. during diastole when the LV reaches its maximum volume. We assigned one segmentation mask for each echocardiography video. Consequently, the segmentation mask assigned to a video was the same assigned to each of its frames. Thus, the final dataset that was used to train the 2D CNN contained the video frames as the input samples, and the  segmentation masks as the labels or the output samples.
\newline \indent 
\subsubsection{Spatial windowing: segmentation process}
\indent

\begin{figure}[h!]
\centering
  \includegraphics[width=1\linewidth]{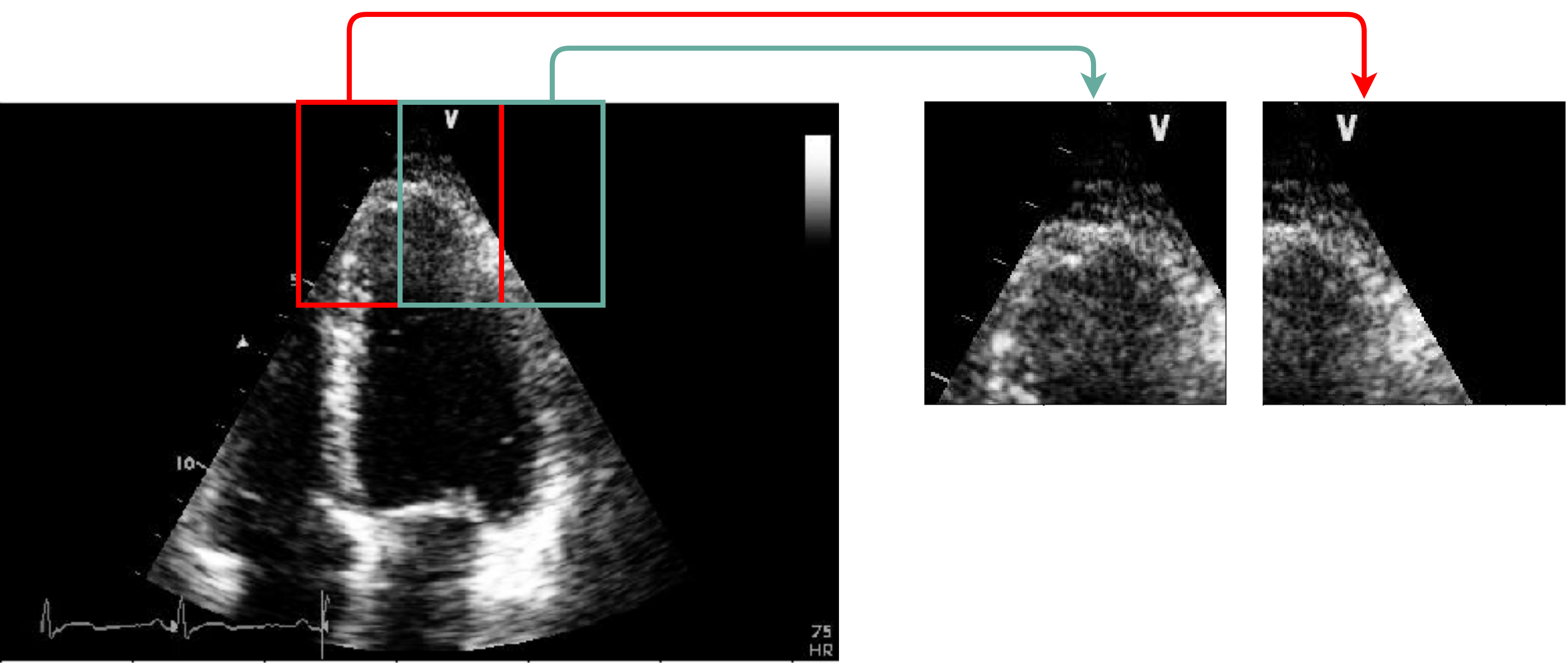}
\caption{Sliding window: the process of extracting two consecutive spatial windows from an frame with an overlap equal to 50\% between the successive windows}
\label{fig:spatial_windowing}
\end{figure}

The next step was to produce frames of the same spatial dimensions (frame size). Thereby, we opted for the sliding window technique to create spatial windows of fixed dimensions, and we applied the technique on both the input samples and the labels. The technique consists of extracting consecutive windows of equal dimensions with an overlap between two successive  windows. Normally, the dimension of the window must be less than or equal to the original dimension of the frame from which it was extracted. Also, the overlap should be less than the dimension of the window. In Figure \ref{fig:spatial_windowing}, we illustrate the sliding window technique, where it extracts two successive windows with an overlap equal to 50\%. The red square in the figure represents a window and the green square represents its successive window that overlaps with the red square by 50\%.
\newline \indent
By applying the sliding window technique on the dataset, we created windows of dimension equal to $150\times150$ px, with a 50\% spatial overlap equal to 75 px. The dimensions of the windows are always less than the original dimensions of the video frames, where the smallest frame dimension in the input samples is equal to $422\times636$ px. In this manner, we obtained a larger and uniform set of input samples totalling 108,127 windows.

The 2D CNN generates an estimation of a segmentation mask for an input window where each value within the segmentation mask is in the interval $[0,1]$. We round these values to obtain a perfect mask with pixel values equal to either 0 or 1. Once the segmentation mask corresponding to each window is estimated, the complete segmentation mask of a video frame is reconstructed using the inverse sliding window technique. The technique is performed by adding the successive estimated segmentation masks of every window from a certain frame with an overlap equal to 50\% until we recover the entire frame. The reconstructed frame has the same dimension as the original frame cut from its video. With the same inverse sliding window technique, we recover all the segmentation video frames and also all the segmentation masks, where each mask corresponds to a frame. Then, having all the segmentation masks predicted for each frame of a given video, we aggregate these masks employing statistical mode (i.e. the most represented value in each pixel is chosen) to form the segmentation mask corresponding to the totality of a video. 

\begin{figure}[h!]
    \centering
    \begin{subfigure}[b]{0.49\textwidth}
        \includegraphics[width=1\linewidth]{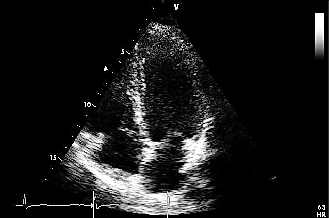}
        \caption{}
        \label{segmentation_a}
    \end{subfigure}
    \begin{subfigure}[b]{0.49\textwidth}
        \includegraphics[width=1\linewidth]{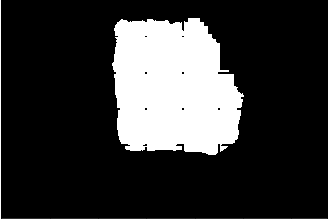}
        \caption{}
        \label{segmentation_b}
    \end{subfigure}
    \begin{subfigure}[b]{0.49\textwidth}
        \includegraphics[width=1\linewidth]{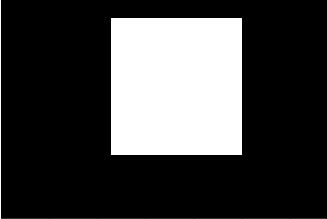}
        \caption{}
        \label{segmentation_c}
    \end{subfigure}
    \begin{subfigure}[b]{0.49\textwidth}
        \includegraphics[width=1\linewidth]{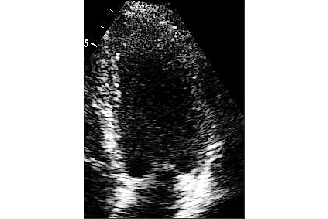}
        \caption{}
        \label{segmentation_d}
    \end{subfigure}
    \caption[]{Illustration of the input and output images from the different phases of the segmentation process performed by a 2D CNN, such that (5a) represents the 2D CNN input image,  and (5b) to (5d) represent the different output images per phase: (5a) represents a video frame captured from an echocardiography video, it also represents the 2D CNN input image. (5b) represents the predicted segmentation mask corresponding to the input frame (5a). It also represents the output image of the 2D CNN. (5c) represents the minimum bounding box estimated from the predicted segmentation mask (5b), and (5d) represents the segmented video frame produced by multiplying the minimum bounding box (5c) by the video frame (5a). }
    \label{fig:5}
\end{figure}

Figure \ref{fig:5} encapsulates the process of applying the predicted segmentation mask on a video frame. Figure \ref{segmentation_a} shows the original video frame, while Figure \ref{segmentation_b} shows its corresponding predicted mask recovered from the reverse sliding window technique, which appears as a set of points with undefined boundaries. Hence, to recover a rectangular-shaped segmentation mask we apply the minimum bounding box technique to enclose the estimated set of points into a rectangle and to produce a bounding box as shown in Figure \ref{segmentation_c}. Then, each video frame is multiplied by its corresponding bounding box to produce a segmented frame as shown in Figure \ref{segmentation_d}. The segmented frames belonging to the same video are then reassembled to produce a segmented video, where the order of appearance of each segmented frame is kept in its same order of appearance as in the original video. The segmented video has the same number of frames as the original video prior to any preprocessing, however, it has fewer frame sizes.  

\subsection{2D CNN Architecture}
\indent

\begin{figure*}[h!]
\centering
  \includegraphics[width=1\linewidth]{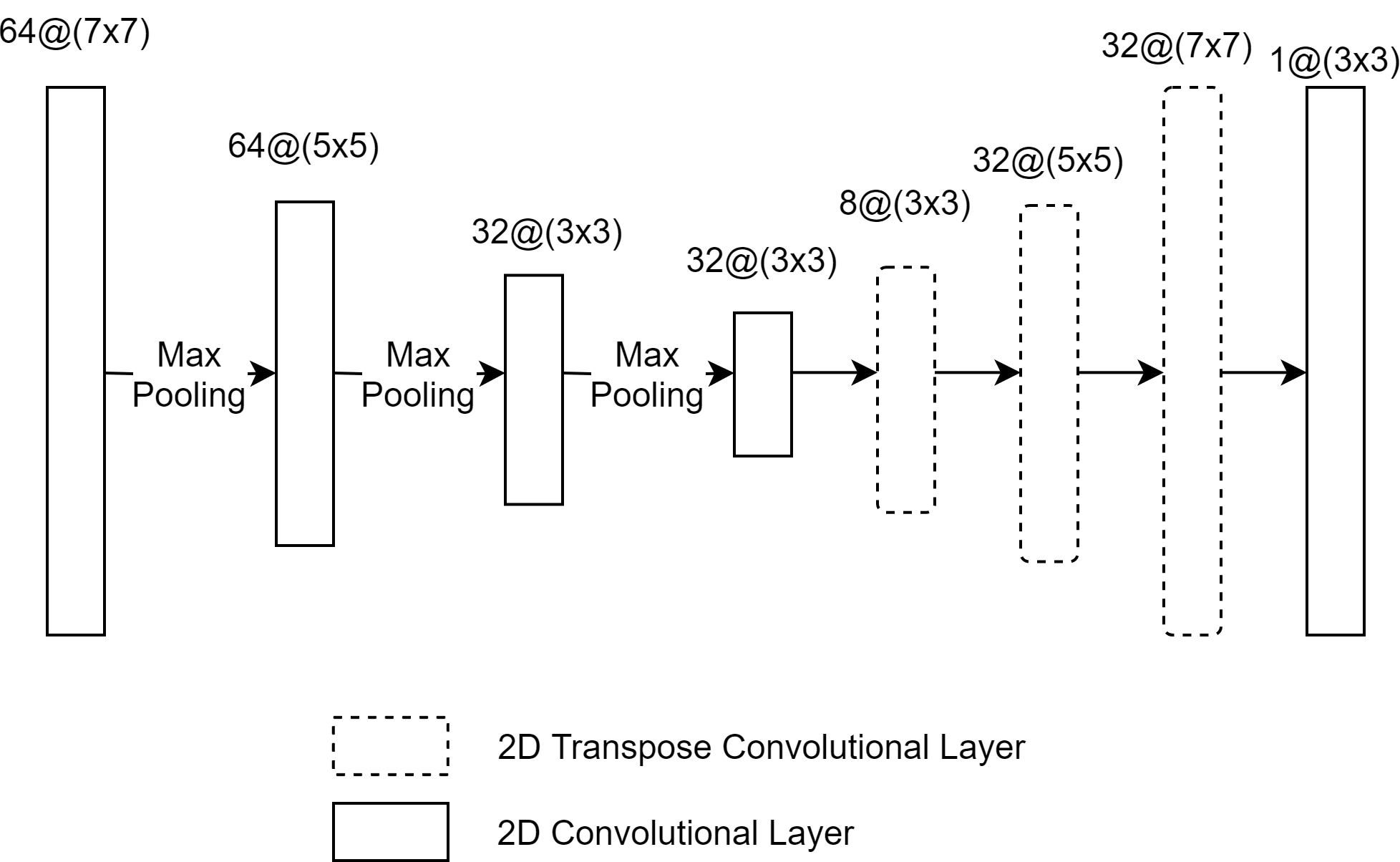}
\caption{The architecture of the 2D CNN}
\label{fig:2d architecte}
\end{figure*}

Our 2D CNN architecture follows the encoder-decoder design common to CNNs developed for semantic segmentation problems \cite{semantic,semantic2,chris}. Figure \ref{fig:2d architecte} illustrates the detailed configuration of the 2D CNN consisting of 3 convolutional layers with rectified linear unit (ReLU) as the activation function for each layer. Every convolutional layer is followed by a max-pooling layer to reduce the dimension of the window. Then, the convolutional layers are followed by 3 transpose convolutional layers \cite{transpose} with a stride equal to $2\times2$ in order to reacquire the initial input dimension. Each transpose layer uses a ReLU as its activation function. 
The last layer is a convolutional layer with a sigmoid activation function, which was selected to produce a predicted segmentation mask with pixel values equal to probabilities between the range of $[0,1]$. The input and output dimensions are $150\times150$ px, which correspond to a segmentation mask adequate for the input window.

\section{MI Detection with 3D CNN}
\label{3dcnn}
\indent

In this section, we give details of our proposed MI detection method using a 3D CNN over the segmented echocardiography videos obtained from a 2D CNN. However, these segmented videos have a different number of frames and different dimensions. In the following section, we give preprocessing details of segmented videos.

\subsection{Data Preprocessing with 3D CNN}
\indent

\begin{figure}[h!]
\centering
  \includegraphics[width=1\linewidth]{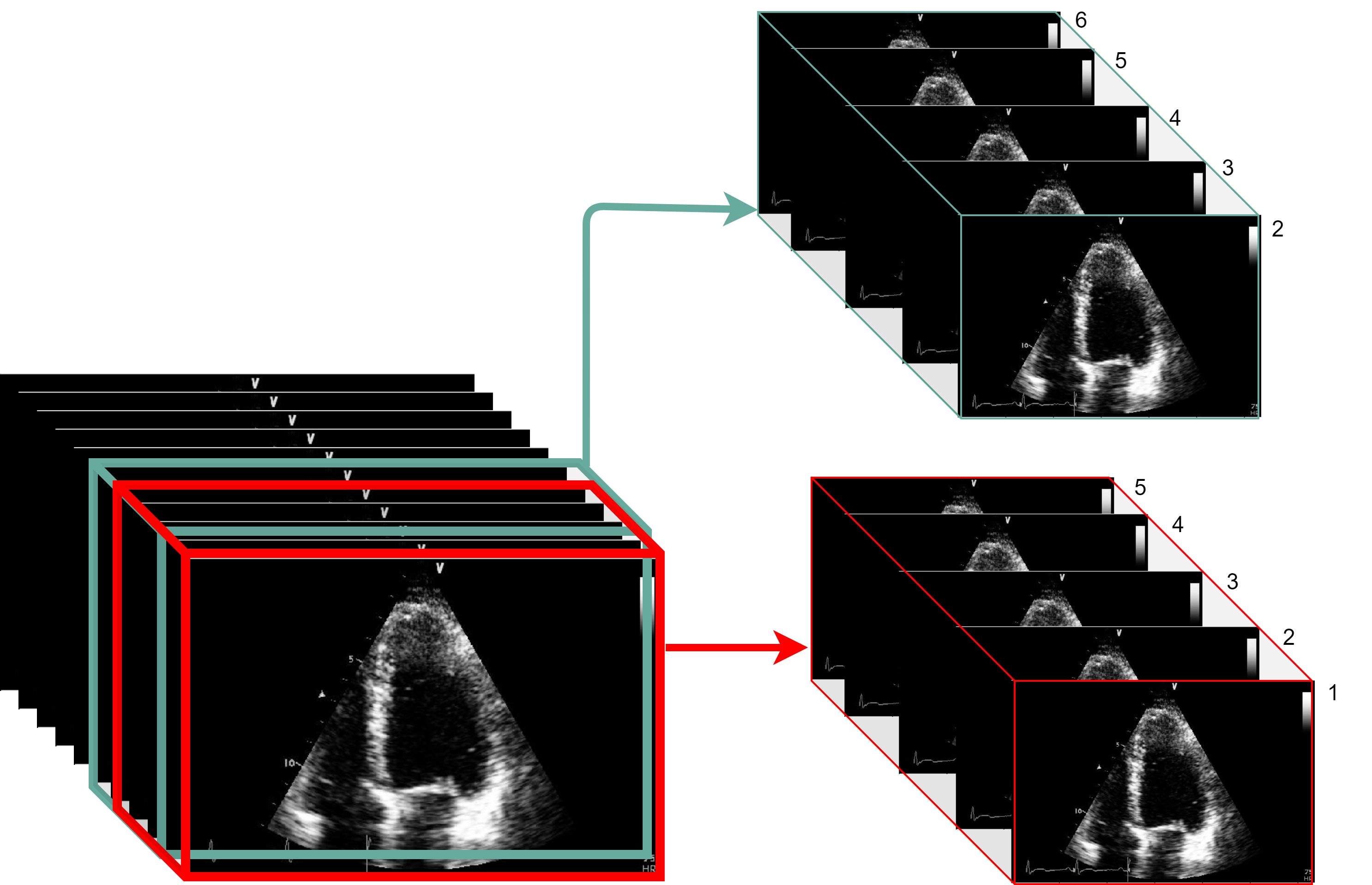}
\caption{Temporal sliding window depicting the process of extracting 2 consecutive temporal windows of size 5 frames, with an overlap equal to 4 frames between two consecutive windows}
\label{fig:temporal_windowing}
\end{figure}

To solve the issue of differences in the spatial dimensions, all the video frames were scaled down to the smallest video size in the dataset. In our case, the smallest frame size from the segmented videos is equal to $236\times183$ px. Then, we applied the sliding window technique to the resized videos in order to obtain a uniform number of frames. The technique consists of extracting a temporal window created from a consecutive number of frames from a given video and repeating the process by going over all the video frames with respect to an overlap between two successive temporal windows. In general, the overlap size is smaller than the temporal window size. The technique allows dividing the dataset videos into smaller temporal windows of a fixed number of frames. It also allowed us to increase the number of samples for the 3D CNN from 165 segmented videos to 2000 temporal windows. In our case, we applied the sliding window technique to extract temporal windows of size equal to 5, 7, and 9 frames per window, with an overlap equal to 4, 6, and 8 frames respectively (i.e. the sliding window moves forward by one window per step). By varying the size of the temporal windows, we created 3 different datasets that we used to train 3 different 3D CNN models.
\newline \indent
We illustrate in Figure \ref{fig:temporal_windowing} the sliding window technique for a temporal window size equal to 5. The red window represents a temporal window consisting of 5 successive frames. The green window is the successive temporal window of the red one that also contains 5 frames, such that the first 4 frames from the green window are the same as the last 4 frames from the red window. 
The labels attributed to these temporal windows are the same as the labels of the video from which these windows were extracted. 

Table \ref{tab:1} shows the number of the temporal windows obtained from the dataset videos by varying the frame number of the temporal windows. For window sizes equal to 5, 7 and 9, we obtained 2841, 2511, and 2181 temporal windows respectively from the dataset of the segmented videos.

\begin{table}[h!]
\centering
\caption{\ Number of windows obtained by applying the temporal sliding window technique with different window sizes}
\label{tab:1}       
\begin{tabular}{lll}
\hline\noalign{\smallskip}
Size of the temporal window & Number of windows \\
\noalign{\smallskip}\hline\noalign{\smallskip}
 5 frames &  $2841$    \\ 
 7 frames &  $2511$    \\
 9 frames &  $2181$   \\
\noalign{\smallskip}\hline
\end{tabular}
\end{table}

In another experiment, we applied a sliding window technique that extracts spatio-temporal windows from the segmented videos in an attempt to avoid rescaling the videos to the smallest dimension. The technique consists of combining the temporal and spatial sliding window techniques at the same time. Even though this  process resulted in a larger dataset, the predicted accuracies were lower than those obtained by simply resizing the segmented videos and applying only temporal sliding window. Therefore, we concluded that the LV chamber should be fully preserved as a frame in the echocardiography video for the 3D CNN to capture all the details throughout the process of learning. Cutting the LV chamber from a segmented video by a spatial sliding window will degrade the information and will result in a poor model.

\subsection{3D CNN Architectures}
\indent

\begin{figure*}[h!]
\centering
  \includegraphics[width=1\linewidth]{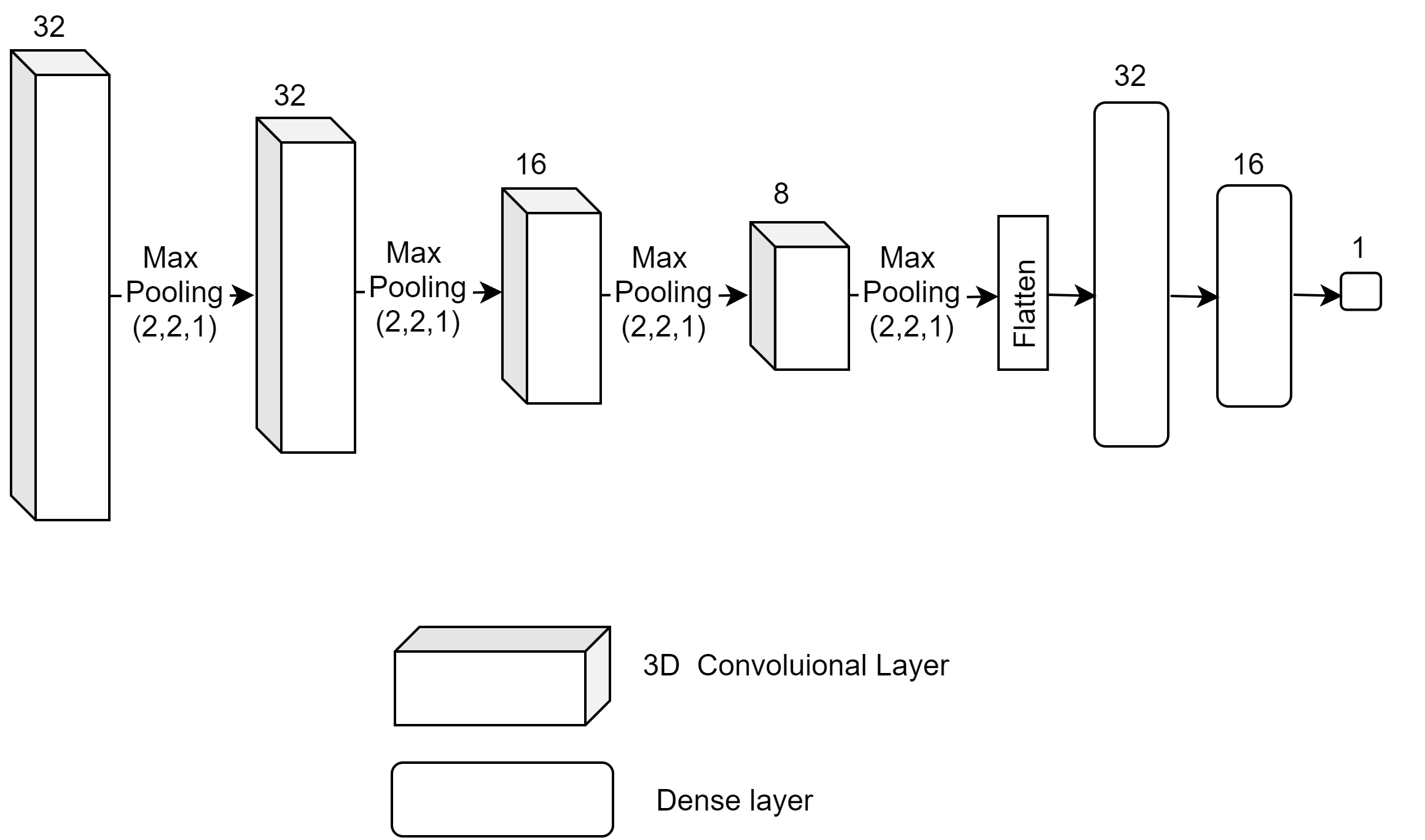}
\caption{The generic architecture of the 3D CNN used to train all the datasets}
\label{fig:3d_architecture}
\end{figure*}

In this section, we present the architectures of the 3D CNN models used to train the 3 datasets separately. For each dataset, we used the same model architecture: same number of layers, the same number of neurons and same activation functions. However, we changed the kernel size for each model to make it fit with the input dimension of the windows. 
\newline \indent
Figure \ref{fig:3d_architecture} shows the architecture of the 3D CNN consisting of 4 3D convolutional layers, 4 2D max-pooling layers, and 3 dense layers. The same activation function was used for all the layers, both convolutional and dense, except for the output layer, is ReLU. For the output layer, which consists of one neuron that contains the prediction probability, we used the sigmoid activation function. Table \ref{tab:2} gives the details of the characteristics of each 3D CNN model.
%
%
\begin{table}[h!]
\centering
\caption{\ 3D CNN characteristics per layer according to the size of the temporal window}
\label{tab:2}
\begin{tabular}{lllll}
\hline\noalign{\smallskip}
 & &  \multicolumn{3}{c}{Kernel size per window size} \\
\noalign{\smallskip}\cline{3-5}\noalign{\smallskip}
Layer & No. of neurons & 5& 7 &9  \\
\noalign{\smallskip}\hline\noalign{\smallskip} 
 Conv3D   &   $32$  & $(3,3,3)$ & $(3,3,3)$& $(3,3,3)$ \\
 MaxPooling   &   - & $(2,2,1)$ & $(2,2,1)$& $(2,2,1)$ \\
 Conv3D   &   $32$ & $(3,3,2)$ & $(3,3,3)$& $(3,3,3)$  \\
 MaxPooling   &  - & $(2,2,1)$ & $(2,2,1)$& $(2,2,1)$ \\
 Conv3D   &   $16$  & $(3,3,2)$ & $(3,3,2)$& $(3,3,3)$  \\
 MaxPooling   &  - & $(2,2,1)$ & $(2,2,1)$& $(2,2,1)$ \\
 Conv3D &   $8$& $(3,3,1)$ & $(3,3,2)$& $(3,3,3)$\\
 MaxPooling   &  - & $(2,2,1)$ & $(2,2,1)$& $(2,2,1)$ \\
 Flatten   &  - & &  &  \\
 Dense &  $32$& & & \\
 Dense & $16$& & & \\
 Dense &  $1$& & & \\
\noalign{\smallskip}\hline
\end{tabular}
\end{table}

\section{Experiments and Results}
\label{results}
\subsection{2D CNN: Training, Evaluation Metrics and Results}
\indent

In order to find the most efficient architecture for the 2D CNN that performs the best on the task of video segmentation, we performed a grid search on the number of layers of the encoder part and the number of neurons per layer. The defined sets of values that were tested for the grid search are $\{3, 4, 5\}$ layers for the encoder and $\{8, 16, 32, 64\}$ neurons per layer. Then, we performed 5-fold cross-validation (CV) \cite{CrossValidation} for each architecture set and selected the one that performed the best amongst all 2D CNN architectures which is described in Figure \ref{fig:2d architecte}. 80\% of the dataset was used for training and 20\% for testing. Since 5-fold CV was used, the training set was further divided into 80\% for training and 20\% for validation. The network that performs best on the validation set was retrained on the whole training set and finally evaluated on the test set. Prior to training, we normalized the data to values in the interval $[0,1]$. We trained all networks for 100 epochs with a batch size equal to 256 resulting in 192,617 trainable parameters for the best 2D CNN. We used the sigmoid activation function for the last layer, and RMSProp optimizer as the optimization function for the 2D CNN. To evaluate the model’s performance we used the mean squared error (MSE) as the loss function. As a result, the MSE is defined as 
\begin{equation}
\label{eq:mse}
\begin{aligned}
MSE(w,\hat{w}) = \frac{1}{w_hw_w}\sum_{i=0}^{w_h-1}\sum_{j=0}^{w_w-1} [w_{(i,j)}-\hat{w}_{(i,j)}]^2.
\end{aligned}
\end{equation}
In Eq. \ref{eq:mse}, $w_h$ and $w_w$ are the window's height and width, respectively, while $w$ and $\hat{w}$ are the actual window and its corresponding prediction, respectively.
\indent
All relevant details regarding the training parameters of the 2D CNN are presented in Table \ref{tab:3}.

\begin{table}[h!]
\centering
\caption{\ 2D CNN training parameters}
\label{tab:3}       
\begin{tabular}{ll}
\hline\noalign{\smallskip}
Parameters & Values  \\
\noalign{\smallskip}\hline\noalign{\smallskip}
 Input samples (windows)   & $86,502 \ (80\%)$    \\ 
 Input shape   & $(150,150)$  \\
 Output shape  &  $(150,150)$  \\
 Trainable parameters   & $192,617$    \\
 Loss&   MSE \\
 Optimizer& RMSProp \\
 Epochs   & $100$ \\
 Batch size&   $256$ \\
\noalign{\smallskip}\hline
\end{tabular}
\end{table}

\subsection{2D CNN: Results and Discussion}
\indent

We evaluated the model using the test set by calculating the accuracy as
\begin{equation}
\label{eq:accuracy}
\begin{aligned}
\ \ \ \ \ \ \ \ \ \ Accuracy(w,\hat{w}) = 1-MSE(w,\hat{w}).
\end{aligned}
\end{equation}
The model achieved 97.18\% accuracy over the test set, which demonstrates that extracting the LV region manually can be replaced by an automatic segmentation method of high precision.

\subsection{3D CNNs: Training and Evaluation Metrics}
\indent

For the 3D CNN experiments, we performed a grid search on the number of layers and the number of neurons per layer for both the convolution layers and the classifier to find a good performing 3D CNN architecture for MI detection in echocardiography. The defined sets of values that were tested for the grid search were $\{3, 4, 5, 6\}$ layers for convolution, $\{2, 3, 4\}$ layers for the classifier, and $\{8, 16, 32, 64\}$ neurons per layer. Then, we performed 5-fold CV for each architecture set and selected the one that performed the best amongst all 3D CNN architecture sets. Prior to training, we normalized the data to values in the interval $[0,1]$. The selected architecture was used for all the 3D CNN models. 
For training all three models, dataset was split into a training set and a test set consisting of 80\% and 20\% of the dataset, respectively. Then, the training set was further split into 80\% for training and 20\% for validation for the CV experiments. Since MI detection is a binary classification, we ensured that the dataset is balanced with respect to N and MI classes. Next, we applied 5-fold CV on each training and validation sets. We retrained the models with the best architectures on the entire training set and we evaluated them on the test set. However, our goal is to predict the class of a complete echocardiography video rather than the class of a temporal window. Thus, to calculate the evaluation metrics of the 3D model over the task of MI detection per video, we assigned a prediction class to each video as the result of the statistical mode calculated over all the predicted classes of the windows constituting that video. The evaluation metrics used to assess the performance of the models are precision, recall (sensitivity), and F1 score.
\newline \indent
To train the models, we used the same loss function and optimizer, however, the input shape varies between the models as shown in Table \ref{tab:4}. We used binary cross-entropy as the loss function \cite{BCE}, and the RMSProp optimizer for all three models. Using the RMSProp optimizer is equivalent to using an adaptive learning rate, therefore, we did not include the learning rate in the set of tunable hyperparameters and we chose $1e^{-3}$ as the initial learning rate. For each fold, we trained the model for 100 epochs using a batch size equal to 8 samples. We calculated the evaluation metrics per video for each fold associated with each model. 
\newline \indent
To implement the 3D CNN models, we used the Python programming language and its open-source neural network library Keras \cite{keras}. We conducted the experiments on a NVIDIA Tesla P100 GPU server with 12GB of GPU memory.

\begin{table}[h!]
\caption{\ 3D CNN models' training parameters per window size}
\label{tab:4}
\centering
\begin{tabular}{llll}
\hline\noalign{\smallskip}
 &  \multicolumn{3}{c}{Window size} \\
\noalign{\smallskip}\cline{2-4}\noalign{\smallskip}
Parameters & 5 & 7 & 9  \\
\noalign{\smallskip}\hline\noalign{\smallskip}
 Input samples (windows)  &  $2273$ & $2009$ & $1745$ \\
 Input shape   &(236,183,5)&(236,183,7)&(236,183,9) \\
 Trainable parameters   &$57,977$ & $68,345$ & $74,105$   \\
  Learning rate  & {$1e^{-3}$} & {$1e^{-3}$} & {$1e^{-3}$}\\
 Loss   & \multicolumn{3}{c}{Binary CrossEntropy} \\
 Optimizer  &   \multicolumn{3}{c}{RMSProp}\\
 Epochs per fold & \multicolumn{3}{c}{$100$}\\
 Batch size  &  \multicolumn{3}{c}{$8$}\\
 \hline 
\end{tabular}
\end{table}

\subsection{3D CNNs: Results and Discussion}
\indent

Table \ref{table:3dresults} shows the results of the evaluation metrics, as produced by the fully trained 3D models using 5-fold CV and calculated with their corresponding test sets. Only the highest, lowest, and mean values are given.

The best results of our models were: 90.9\% accuracy, 97.2\% F1 score, 100\% precision, and 95\% recall. However, the mean values for the evaluation metrics are slightly lower than the maximum values, and this is explained by the fact that the training sets contain distinct training samples, where some of the windows contain more noise and hence are of poorer quality than other windows. Therefore, even though all the sets contain balanced and equal proportions of samples representing both classification classes, some of the folds may contain more noisy samples than the remaining folds, which influences the learning performance of the model at each fold. Hence, the model trained over the dataset of windows with the size equal to 5 frames, achieved the following mean values over the 5 folds of CV experiments: 84.6\% accuracy, 87\% F1 score, 89\% precision, and 85.1\% recall. Furthermore, we observe that the mean values of the evaluation metrics obtained from the dataset of windows with a size equal to 7 frames are slightly inferior to those attained from the dataset of windows with a size equal to 5. Likewise, the mean values of the evaluation metrics achieved over the dataset of windows equal to 9 frames, are less than those obtained over the windows of size equal to 7 frames. The mean values of the metrics obtained from the second dataset (window size 7) are: 82.5\% accuracy, 83.2\% F1 score, 83.5\% precision and 83.1\% recall, whereas, the values obtained from the third (window size 9) are: 81.3\% accuracy, 83.1\% F1 score, 84.6\% precision and 82\% recall. Thus, we conclude that enlarging the size of the temporal window reduces the performance of the 3D CNN. On an average, the end-to-end system lasts $13.12$ milliseconds to predict MI per video.

\begin{table*}[h!]
\caption{\ 3D CNN models' evaluation metrics per window size}
\label{table:3dresults}
\centering
\begin{tabular}{lllll}
\hline\noalign{\smallskip}
 &  &\multicolumn{3}{c}{Window size} \\
\noalign{\smallskip}\cline{3-5}\noalign{\smallskip}
Evaluation metrics & & 5 & 7 & 9  \\
\noalign{\smallskip}\hline\noalign{\smallskip} 

 \multirow{3}{*}{Accuracy}& Max & $90.3 \ \%$ & $\textbf{90.9 \%}$ & $90.0\ \%$ \\
 \cline{2-5}
 \multirow{3}{*}{}  & Mean &$\textbf{84.6\ \%}$& $82.5\ \%$ & $81.3\ \%$ \\
 \cline{2-5}
 \multirow{3}{*}{}& Min & $\textbf{77.1\ \%}$ & $72.9\ \%$ & $68.4\ \%$ \\
 
 \hline
 \multirow{3}{*}{F1 score}& Max &$94.8\ \%$ & $\textbf{97.2\ \%}$ & $92.3\ \%$ \\
 \cline{2-5}
 \multirow{3}{*}{}  &  Mean &$\textbf{87.0\ \%}$ & $83.2 \ \%$ & $83.1\ \%$ \\
 \cline{2-5}
 \multirow{3}{*}{}&  Min &$68.7\ \%$ & $\textbf{75.0 \ \%}$ & $68.4\ \%$ \\
 
 \hline
 \multirow{3}{*}{Precision}& Max &$94.7\ \%$ & $\textbf{100\ \%}$ & $94.7\ \%$ \\
 \cline{2-5}
 \multirow{3}{*}{}  & Mean & $\textbf{89.0\ \%}$ & $83.5 \ \%$ & $84.6\ \%$ \\
 \cline{2-5}
 \multirow{3}{*}{}& Min &$73.0\ \%$ & $\textbf{75.0\ \%}$ & $72.2\ \% $ \\
 \hline
 \multirow{3}{*}{Recall}& Max & $\textbf{95.0\ \%}$ & $94.7 \ \%$ & $90.0\ \%$ \\
 \cline{2-5}
 \multirow{3}{*}{}  & Mean &$\textbf{85.1\ \%}$ & $83.1\ \%$ & $82.0\ \%$ \\
 \cline{2-5}
 \multirow{3}{*}{}& Min &$65.0\ \%$ & $\textbf{75.0\ \%}$ & $65.0\ \%$ \\
 \hline
\end{tabular}
\end{table*}

Table \ref{table:comparison} represents a performance comparison between the results of our 3D CNN models on MI detection in echocardiography and two recent state-of-the-art works \cite{qatar3,qatar1}. Both the results of our model and the state-of-the-art works were achieved on the HMC-QU dataset. \cite{qatar3} used an Active Polynomial-based model to detect MI in the HMC-QU dataset and reported the results from two separate tests. The first test was conducted on the HMC-QU dataset, while the second test was conducted over a partition of the dataset that included only videos with reasonable quality. \cite{qatar1} used an E-D CNN to achieve the MI detection in the HMC-QU dataset. To report the classification results, two types of experiments were conducted. In the first type, they extracted 6-segment features form the echocardiography and used them to train four conventional classification models: linear discriminant analysis (LDA), decision tree (DT), random forest (RF), and support vector machine (SVM). While in the second type of experiment, they extracted 5-segment features and used them to train the same four classifiers. 
In Table \ref{table:comparison}, we report the results of our three 3D CNN models, in addition to the results of Active Polynomials model tested over both reasonable quality videos and the entirety of the HMC-QU dataset, and the results of the four classifiers of the E-D CNN tested over both 6-segment features data and 5-segment features data. 
\newline \indent
The values of the performance metrics show the significant superiority of our 3D CNN MI prediction model in terms of accuracy, F1 score, precision and recall over state-of-the-art models on the HMC-QU dataset. Even compared to the results obtained by the Active Polynomials model tested only on reasonable-quality videos, and which has the highest performance metrics amongst the remaining state-of-the-art methods, our 3D CNN achieved remarkably higher results of accuracy, F1 score, precision and recall over the entirety of the HMC-QU dataset that contains both low-quality and reasonable-quality videos. The Active Polynomials model tested over reasonable quality videos only achieved 87.9\%, 90.1\%, 87.6\%, and 92.8\% of accuracy, F1 score, precision and recall respectively, while the 3D CNN achieved 90.9\%, 97.2\%, 100\%, and 95\% of accuracy, F1 score, precision and recall respectively, which shows the robustness and the efficiency of the 3D CNN. Active Polynomials tested on the HMC-QU dataset achieved slightly better results compared to the remaining methods, followed by the SVM model with 5-segment features, however, their corresponding results are still lower than those achieved by the 3D CNN.  

\begin{table*}[h!]
\caption{\ Performance comparison of the results from the 3D CNN model presented in this paper with the state-of-the-art methods: Active Polynomials \cite{qatar3} and E-D CNN \cite{qatar1}. All MI detection results in this table were produced using the HMC-QU dataset}
\label{table:comparison}
\centering
\begin{tabular}{llllll}
\hline\noalign{\smallskip}
 & \multicolumn{3}{p{3cm}}{\centering 3D CNN} &\multicolumn{2}{p{3.25cm}}{\centering Active Polynomials \cite{qatar3}}  \\

 \noalign{\smallskip}\cline{2-6}\noalign{\smallskip}
 & \multicolumn{3}{p{3cm}}{\centering Window size}   & \multicolumn{2}{r}{\centering Video quality\ \ \ \ }  \\
 
  \noalign{\smallskip}\cline{2-6}\noalign{\smallskip}
\multicolumn{1}{p{3cm}}{Evaluation metrics}& \multicolumn{1}{l}{\centering 5} & \multicolumn{1}{l}{\centering 7} & \multicolumn{1}{l}{\centering 9} & \multicolumn{1}{r}{\ \ \ \ Reasonable} & \multicolumn{1}{r}{\centering All \ \ } \\

\noalign{\smallskip}\cline{1-6}\noalign{\smallskip}
\multicolumn{1}{p{1.25cm}}{Accuracy}	&\multicolumn{1}{l}{\centering $90.3\%$}	&\multicolumn{1}{l}{\centering \textbf{90.9\%}}	&\multicolumn{1}{p{0.75cm}}{\centering 90.0\%}	&\multicolumn{1}{r}{\centering 87.9\%}	&\multicolumn{1}{r}{\centering 83.1\%}\\	

\noalign{\smallskip}\cline{1-6}\noalign{\smallskip}
\multicolumn{1}{p{1.25cm}}{F1 score} &	\multicolumn{1}{p{0.75cm}}{\centering 94.8\%} &	\multicolumn{1}{p{0.75cm}}{\centering \textbf{97.2\%}}&	\multicolumn{1}{p{0.75cm}}{\centering 92.3\%}&	\multicolumn{1}{r}{\centering 90.1\%}&\multicolumn{1}{r}{\centering 85.7\%}\\

\noalign{\smallskip}\cline{1-6}\noalign{\smallskip}
\multicolumn{1}{p{1.25cm}}{Precision}&\multicolumn{1}{p{0.75cm}}{\centering	94.7\%}& \multicolumn{1}{p{0.75cm}}{\centering \textbf{100\%}}& \multicolumn{1}{p{0.75cm}}{\centering 94.7\%}& \multicolumn{1}{r}{\centering	87.6\%}& \multicolumn{1}{r}{\centering	82.6\%}\\
\noalign{\smallskip}\cline{1-6}\noalign{\smallskip}

\multicolumn{1}{p{1.25cm}}{Recall} & \multicolumn{1}{p{0.75cm}}{\centering \textbf{95.0\%}}& \multicolumn{1}{p{0.75cm}}{\centering 94.7\%} & \multicolumn{1}{p{0.75cm}}{\centering 90.0\%}& \multicolumn{1}{r}{\centering 92.8\%} & \multicolumn{1}{r}{\centering 89.0\%}\\
\noalign{\smallskip}\cline{1-6}\noalign{\smallskip}
\end{tabular}

\begin{tabular}{lllllllll}
\multicolumn{3}{p{4cm}}{} & \multicolumn{6}{p{3.75cm}}{\centering E-D CNN \cite{qatar1}}   \\
\noalign{\smallskip}\cline{2-9}\noalign{\smallskip}
 
  & \multicolumn{4}{p{3cm}}{\centering 6-segment features}   & \multicolumn{4}{p{3cm}}{\centering \ \ \ \ \ \ 5-segment features}  \\
   \noalign{\smallskip}\cline{2-9}\noalign{\smallskip}
   
 \multicolumn{1}{p{3cm}}{Evaluation metrics}  & \multicolumn{1}{l}{LDA}   & \multicolumn{1}{p{0cm}}{DT}   & \multicolumn{1}{p{0.5cm}}{RF}  & \multicolumn{1}{l}{SVM} & \multicolumn{1}{l}{LDA}  & \multicolumn{1}{l}{DT}   & \multicolumn{1}{p{0.5cm}}{RF}  & \multicolumn{1}{p{0.5cm}}{SVM}  \\
 \noalign{\smallskip}\cline{1-9}\noalign{\smallskip}
  
 \multicolumn{1}{p{3cm}}{ Accuracy} & \multicolumn{1}{p{0.5cm}}{\centering 75.6\%}  & 
\multicolumn{1}{p{0.5cm}}{\centering 72.6\%} & \multicolumn{1}{p{0.5cm}}{\centering 77.4\%} &\multicolumn{1}{p{0.5cm}}{\centering 80.2\%} & \multicolumn{1}{p{0.5cm}}{\centering 78.5\%} & \multicolumn{1}{p{0.5cm}}{\centering 73.5\%} & \multicolumn{1}{p{0.5cm}}{\centering 76.6\%} & \multicolumn{1}{p{0.5cm}}{\centering 80.2\%} \\
 \noalign{\smallskip}\cline{1-9}\noalign{\smallskip}

\multicolumn{1}{p{3cm}}{F1 score} & \multicolumn{1}{p{0.5cm}}{\centering 80.6\%}  & 
\multicolumn{1}{p{0.5cm}}{\centering 79.4\%} & \multicolumn{1}{p{0.5cm}}{\centering 82.5\%} &\multicolumn{1}{p{0.5cm}}{\centering 85.2\%} & \multicolumn{1}{p{0.5cm}}{\centering 83.2\%} & \multicolumn{1}{p{0.5cm}}{\centering 80.0\%} & \multicolumn{1}{p{0.5cm}}{\centering 82.6\%} & \multicolumn{1}{p{0.5cm}}{\centering 84.8\%} \\
\noalign{\smallskip}\cline{1-9}\noalign{\smallskip}

\multicolumn{1}{p{3cm}}{Precision} & \multicolumn{1}{p{0.5cm}}{\centering 83.8\%}  & 
\multicolumn{1}{p{0.5cm}}{\centering 80.4\%} & \multicolumn{1}{p{0.5cm}}{\centering 85.9\%} &\multicolumn{1}{p{0.5cm}}{\centering 85.5\%} & \multicolumn{1}{p{0.5cm}}{\centering 86.6\%} & \multicolumn{1}{p{0.5cm}}{\centering 81.7\%} & \multicolumn{1}{p{0.5cm}}{\centering 84.9\%} & \multicolumn{1}{p{0.5cm}}{\centering 86.8\%} \\

\noalign{\smallskip}\cline{1-9}\noalign{\smallskip}
\multicolumn{1}{p{3cm}}{Recall} & \multicolumn{1}{p{0.5cm}}{\centering 78.5\%}  & 
\multicolumn{1}{p{0.5cm}}{\centering 79.0\%} & \multicolumn{1}{p{0.5cm}}{\centering 80.2\%} &\multicolumn{1}{p{0.5cm}}{\centering 85.9\%} & \multicolumn{1}{p{0.5cm}}{\centering 81.3\%} & \multicolumn{1}{p{0.5cm}}{\centering 79.0\%} & \multicolumn{1}{p{0.5cm}}{\centering 82.2\%} & \multicolumn{1}{p{0.5cm}}{\centering 83.0\%} \\
\noalign{\smallskip}\cline{1-9}\noalign{\smallskip}
\end{tabular}
\end{table*}

\section{Conclusion and Future Work}
\label{conclusion}
\indent

In this paper, we proposed a novel, real-time, and fully automated approach consisting of 2D, and 3D CNN models to early detect MI from the echocardiography of a patient. This approach replaces the time-consuming and manual preprocessing with an automated, fast and reliable LV segmentation; and enhances the accuracy and efficacy of MI detection. This system is indiscriminative, in that it processes echocardiography videos of different sizes, frame rates, and resolutions, and it is lightweight in that it runs on parallel threads and does not require high memory or computational power. 

The proposed 2D CNN model for video segmentation achieved a high accuracy of 97.18\% in segmenting LV from the A4C view, therefore, it could be a very reliable and valuable tool to assist cardiologists. Moreover, the proposed 3D CNN models showed that real-time prediction of MI from a patient's echocardiography is realizable, reliable and efficient by achieving 90.9\% accuracy, 100\% precision, 95\% recall, and 97.2\% F1 score. 
Since the HMC-QU dataset included a mix of videos of poor-quality, low-resolution and corrupted by noise, we believe that these factors had a negative impact on the performance of the 3D CNN models on MI detection from the segmented LV views. Nevertheless, the results showed the robustness and efficiency of the proposed approach. Temporal window size has a significant bearing on the performance of the 3D CNN models as evidenced by our experiments. We related this variability in performance to the difference in the 3D CNN model's characteristics, which may have altered the ability of the model to extract relevant prediction features with the given neuron and layer parameters. The 3D CNN models were built with the objective of assigning the least possible number of layers and neurons while still being able to extract relevant spatio-temporal features from the spatio-temporal windows to achieve the most accurate MI detection. 

For our future work, we plan on improving our model's results by enlarging the dataset with more echocardiography videos \cite{data_size}. We believe that training the models with more examples result in significant improvements in performance accuracy. Moreover, training the prediction models using spatio-temporal windows of higher temporal size could boost the 3D CNN accuracy by enabling it to learn more temporal features. However, this requires more computational power and training time. Furthermore, we aim to merge our end-to-end automated pipeline into an embedded system using TensorRT \cite{online2}.

\section*{Acknowledgements}
\indent

The authors wish to acknowledge the valuable contribution of researchers at Medical Research Centre at Hamad Medical Corporation in the State of Qatar for the creation of this work and this publication.

\section*{Declarations}
\indent

\subsection*{Funding}
\indent

The work of Sheela Ramanna and Christopher J. Henry was funded by the NSERC Discovery Grants Program (nos.
194376, 418413).

\subsection*{Conflicts of interest/Competing interests}
\indent

There is no conflict of interest with the funders.

\subsection*{Consent to participate}
\indent

All the Authors consent to the content of the manuscript.

%

\bibliographystyle{unsrt}
\bibliography{bib}
\end{document}